\documentclass[amsmath,showpacs,aps,prb,twocolumn,superscriptaddress,10pt]{revtex4-2}
\usepackage[utf8]{inputenc}
\usepackage{amssymb}
\usepackage{color}
\usepackage{braket}
\usepackage{graphicx}
\usepackage{dcolumn}
\usepackage{bm}
\usepackage{multirow}
\usepackage{overpic}
\usepackage[colorlinks=true,linkcolor=blue,anchorcolor=blue,citecolor=blue,urlcolor=blue]{hyperref}
\allowdisplaybreaks

\begin{document}

\title{Relationship between doping‑induced in‑gap states and spin excitations in Kitaev‑Hubbard models}

\author{Si-Qi Hou}
\affiliation{National Laboratory of Solid State Microstructures and Department of Physics, Nanjing University, Nanjing 210093, China}

\author{Shun-Li Yu}
\affiliation{National Laboratory of Solid State Microstructures and Department of Physics, Nanjing University, Nanjing 210093, China}
\affiliation{Collaborative Innovation Center of Advanced Microstructures, Nanjing University, Nanjing 210093, China}
\affiliation{Jiangsu Key Laboratory of Quantum Information Science and Technology, Nanjing University, Suzhou 215163, China}

\author{Zhao-Yang Dong}
\email[]{zhydong@njust.edu.cn}
\affiliation{Department of Applied Physics, Nanjing University of Science and Technology, Nanjing 210094, China.}

\author{Jian-Xin Li}
\email[]{jxli@nju.edu.cn}
\affiliation{National Laboratory of Solid State Microstructures and Department of Physics, Nanjing University, Nanjing 210093, China}
\affiliation{Collaborative Innovation Center of Advanced Microstructures, Nanjing University, Nanjing 210093, China}
\affiliation{Jiangsu Key Laboratory of Quantum Information Science and Technology, Nanjing University, Suzhou 215163, China}

\date{\today}

\begin{abstract}
We investigate the connection between doping-induced in-gap states and underlying spin excitations in Mott insulators by employing cluster perturbation theory on one-dimensional (1D) and quasi-1D Kitaev-Hubbard models. By manipulating Kitaev-like hopping terms ($t^{\prime}$) that selectively control spin anisotropies in the strong-coupling limit, we establish a direct correspondence between the kinetic dispersion of the in-gap states and the spin excitation spectra. Specifically, in the Z chain, in-gap states evolve from a gapless dispersion to a gapped flat band as the system transitions from the Heisenberg to the Ising model, exhibiting a gap scaling of $2t^{\prime 2}/U$ that matches the Ising spin gap. In the XY chain, the in-gap states split into a dispersive and a flat branch at the Kitaev limit, perfectly mirroring the Jordan-Wigner fermionic spectrum. For the two-leg ladder, we observe an emergent broad continuum of in-gap states that reflects the fractionalization of spin excitations, accompanied by a gap manifesting the presence of topological $Z_2$ visons. Our results establish a robust correspondence between charge and spin dynamics in doped Mott insulators and demonstrate that in-gap states can serve as a probe of exotic quantum spin phenomena, including fractionalization and topological excitations, offering a new pathway to investigate spin liquids via spectroscopic probes of charge excitations.
\end{abstract}

\maketitle

\section{introduction}
\label{sec1}
The undoped parent compounds of the copper oxide high-temperature superconductors are Mott insulators exhibiting long-range antiferromagnetic order. Introducing charge carriers into these systems suppresses the magnetic order and can induce superconductivity below a transition temperature $T_c$. Consequently, elucidating the evolution of the single-particle spectral function with doping has thus become an important issue in the field of high-temperature superconductivity~\cite{RevModPhys.78.17, dagotto_correlated_1994, PhysRevLett.85.2585, RevModPhys.82.1719}.

Mott insulators are a class of insulators caused by strong electron correlations. Studying Mott physics based on the Hubbard model has become a classic paradigm in this field~\cite{hubbard_electron_1963,RevModPhys.82.1719,RevModPhys.70.1039,kmn8-y59j}. Strong Coulomb interactions open a Mott gap, and doping introduces additional electronic states within this Mott gap~\cite{PhysRevB.48.3916,PhysRevLett.67.1035,cpl_40_8_087101}. These additional states are commonly referred to as in-gap states. The appearance of in-gap states signifies a transfer of spectral weight: the spectral weight of the upper Hubbard band (UHB) and lower Hubbard band (LHB) reduces, with the lost portion transferred into the gap, exactly equating to the spectral weight of the in-gap states~\cite{PhysRevB.48.3916,PhysRevLett.67.1035}. Several theoretical perspectives have been proposed to explain the microscopic origin of the in-gap states. A cluster dynamical mean-field theory based analysis suggests that these states originate from a relaxation of the binding strength between doubly occupied and empty sites~\cite{PhysRevLett.102.056404,Phillips_2009}. The mean-field theory based on holo-electron and doublo-hole implies that the hybridization of these composite fermions with conventional quasiparticles gives rise to pseudogaps, Fermi pockets, and Fermi arcs~\cite{PhysRevLett.106.016404,PhysRevB.83.214522}. Alternatively, the in-gap states may be a spin-polaron shakeoff band~\cite{PhysRevB.54.3576,PhysRevB.83.205137}. Further numerical and theoretical results revealed that the spectral weight of the emergent in-gap states vanishes in the zero-doping limit, while their dispersion closely follows that of spin waves in the one-dimensional (1D) antiferromagnetic Heisenberg model, albeit with a momentum shift of a Fermi wavevector~\cite{kohno_spectral_2010,kohno_mott_2012,PhysRevB.92.085129,Kohno_2018}. The origin is also attribute to the coupling of the emergent states with antiferromagnetic correlations~\cite{PhysRevLett.75.1344}. These previous studies indicate that the properties and microscopic origins of in-gap states are not yet fully resolved. A particularly intriguing question is whether a systematic correspondence exists between in-gap states and spin excitations.  

The Kitaev-Hubbard model extends the Hubbard model by incorporating Kitaev-like hopping terms that depend on the direction of nearest-neighbor (NN) bonds, which has been realized in optical lattice systems of ultracold atoms~\cite{duan_controlling_2003}. First-principles calculations indicate that iridates Na$_2$IrO$_3$ includes similar anisotropic hopping terms~\cite{PhysRevLett.102.256403}. Notably, the combined effect of Kitaev-like hopping and Hubbard repulsion can induce novel quantum states, including various charge and magnetic orders~\cite{PhysRevB.89.115130,PhysRevLett.110.037201,PhysRevB.94.125120,PhysRevLett.114.167201}. Our focus lies on the strong-coupling limit of this electronic model, where the low-energy effective spin model reduces to the Kitaev-Heisenberg model~\cite{PhysRevLett.110.037201}. Specifically, in the absence of Kitaev-like hopping, the spins are coupled by NN antiferromagnetic Heisenberg exchanges, which supports well-defined gapless $S=1$ magnon excitations~\cite{PhysRev.128.2131}; when the strength of Kitaev-like hopping equals that of conventional hopping, the effective model is dominated by anisotropic Kitaev spin interactions. The spin spectrum of Kitaev spin liquid forms a continuum due to fractionalized Majorana fermions and gauge-flux excitations~\cite{kitaev_anyons_2006,PhysRevLett.119.157203}. Despite this extensive knowledge of spin excitations, the doping-induced charge excitations remain largely unexplored, in particularly, the relationship between the two. To address these gaps, we focus on 1D and quasi-1D cases because we can control the gap nature of spin excitations by constructing appropriate 1D Kitaev-Hubbard model and tuning the Kitaev-like hopping, where the presence or absence of an excitation gap provides a qualitative criterion for identifying the relationship between charge and spin excitations.  

In this work, we systematically compute the zero-temperature single-particle spectral functions of hole-doped 1D (or quasi-1D) Kitaev-Hubbard models defined on different spatial lattice and specific bond assignments. We explicitly track the evolution of the emergent charge states by tuning the Kitaev-like hopping amplitude $t^{\prime}$. Furthermore, comparing these charge spectra against the corresponding effective spin dynamics reveals a definitive correspondence between the kinetic dispersion of the doping-induced in-gap states and the underlying spin excitations across three distinct lattice geometries. Crucially, while each of these geometries harbors fundamentally distinct spin dynamics, we demonstrate that their unique characteristic signatures are all imprinted onto the single-particle spectra of the corresponding electronic models.

The paper is organized as follows. In Sec.~\ref{sec2}, we introduce the 1D and quasi-1D Kitaev-Hubbard models and outline our numerical CPT methodology. Section~\ref{sec3} details the numerical results and discusses the relationship between charge and spin excitations for three lattice geometries, respectively. Finally, Sec.~\ref{sec4} provides a summary and perspective on our findings.

\section{Models and Method}
\label{sec2}
The Kitaev-Hubbard model is given by:
\begin{equation}
\begin{aligned}
H=&-\sum_{\langle ij\rangle_\alpha}\left\{c_i^{\dagger}\left(\frac{t+t^{\prime} \sigma_\alpha}{2}\right) c_j+\text{H.c.}\right\}+U\sum_i n_{i \dagger} n_{i \downarrow} \\
  &-\mu \sum_{i, \sigma} n_{i \sigma},
\end{aligned}
\label{kitaevhubbardmodel}
\end{equation}
where $\langle i j\rangle_{\alpha}$ denotes summation over NN bonds, $c_{i \sigma}$ is the fermion annihilation operator for spin $\sigma$ at site $i$ and $c_{i}^{\dagger}=(c_{i \uparrow}^{\dagger}, c_{i \downarrow}^{\dagger})$, $\sigma_{\alpha}$ is the Pauli matrix ($\alpha = x,y,z$, dependent on the bond type). We set the conventional hopping amplitude $t=1$ as the energy unit. $t^{\prime}$ represents the Kitaev-like spin-dependent hopping. $U$ parametrizes the on-site Coulomb repulsion, and $\mu$ is the chemical potential. $\mu = U/2$ in our calculations to show the exact particle-hole symmetry at half-filling. In the strong-coupling limit ($U \gg t, t^{\prime}$), the low-energy effective spin Hamiltonian of the electronic model [Eq.~\ref{kitaevhubbardmodel}] maps onto the extensively studied Kitaev-Heisenberg model:
\begin{equation}
H=\sum_{\langle i j\rangle_{\alpha}} K S_i^\alpha S_j^\alpha+J S_i \cdot S_j,
\label{effhamiltonian}
\end{equation}
where the effective exchange couplings are given by $K=2 t^{\prime 2} /U$ and $J=(1-t^{\prime 2})/U$~\cite{PhysRevLett.110.037201}.

\begin{figure}
\centering
\includegraphics[width=0.48\textwidth]{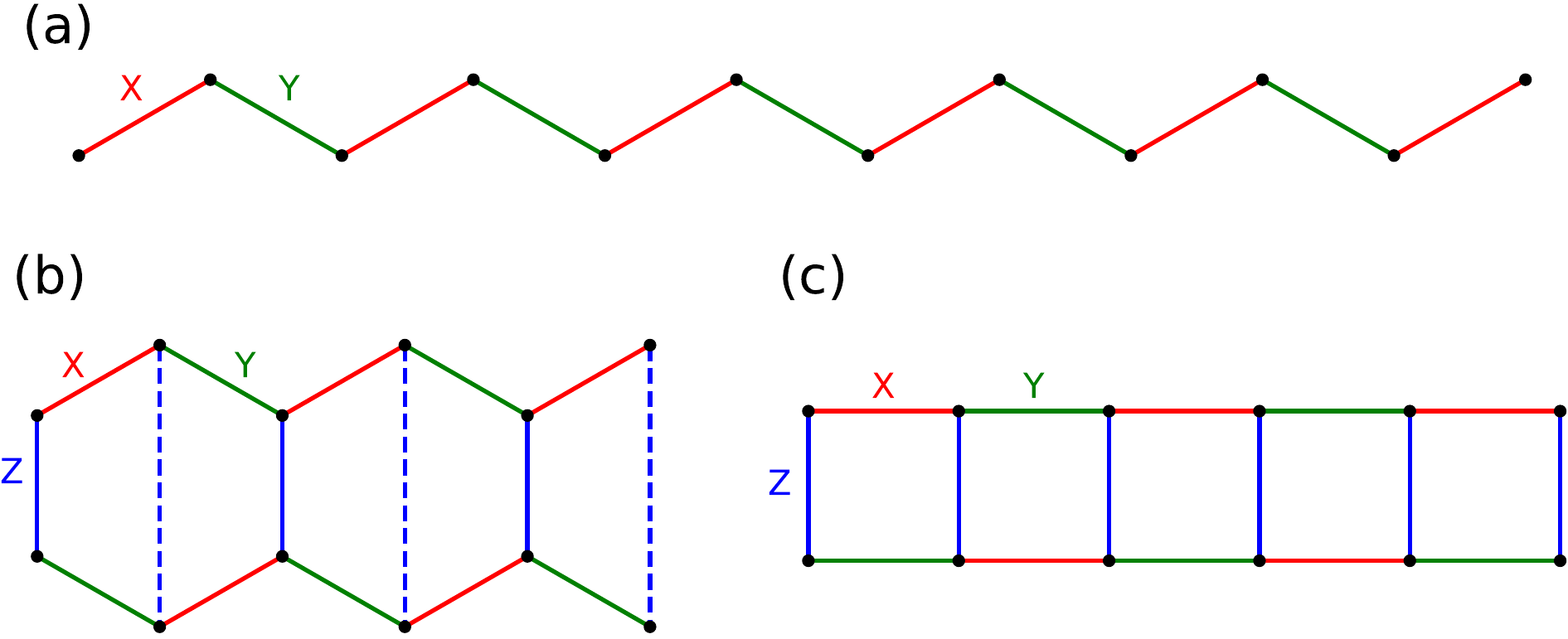}
\caption{12-site clusters used in our calculations. (a) Single chain extracted from the honeycomb lattice. Black dots represent lattice sites, red lines represent $x$-type NN bonds, and green lines represent $y$-type NN bonds. (b) A two-leg ladder extracted from the honeycomb lattice. Both solid blue lines and dashed blue lines represent $z$-type bonds. (c) A two-leg ladder extracted from the square lattice, which is topologically equivalent to (b).}
\label{showclusters}
\end{figure}

In this work, we focus on the 1D limit and specifically investigate three distinct lattice geometries: a 1D Z chain consisting solely of $z$-type NN bonds; a 1D XY chain with alternating $x$-type and $y$-type NN bonds that form a dimerized structure [Fig.~\ref{showclusters}(a)]; and a two-leg ladder extracted from two adjacent chains of the two-dimensional honeycomb lattice [Fig.~\ref{showclusters}(b)]. For the ladder geometry, all three types of NN bonds are present, and we impose periodic boundary condition along the rung direction. Consequently, the ladder includes additional hopping terms (represented by blue dashed lines) that, dictated by the translation symmetry of the ladder, are equivalent to $z$-type bonds. Thus, this configuration is strictly topologically equivalent to a two-leg ladder defined on a square lattice, as illustrated in Fig.~\ref{showclusters}(c).

Let us first consider the Z chain, where $\sigma_{\alpha}$ is restricted to $\sigma_{z}$ for all bonds. In this case, the effective spin model reduces exactly to the 1D spin-1/2 XXZ chain. In the limit $t^{\prime}=0$, it recovers the isotropic 1D antiferromagnetic Heisenberg model. Conversely, at $t^{\prime}=1$, it becomes the 1D antiferromagnetic Ising model, whose ground state features a N\'{e}el order (alternating up and down spins). In this Ising limit, a single spin-flip excitation breaks the lowest-energy configurations of two adjacent bonds, incurring an energy cost of $K$ relative to the ground state. Consequently, as $t^{\prime}$ increases from zero, a finite energy gap is dynamically generated in the originally gapless spin excitation spectrum of the Heisenberg chain, see Appendix~\ref{appe-XXZ}.

Next, we turn to the XY chain and the two-leg ladder configurations extracted from the honeycomb lattice. At the pure Kitaev point ($t^{\prime}=1$), Heisenberg terms disappear and these Kitaev spin models can be mapped exactly onto spinless fermion systems analogous to the 1D $p$-wave BCS superconductors by employing the Jordan-Wigner transformation~\cite{PhysRevLett.98.087204,degottardi_topological_2011,Chen_2008}. The detailed analytical derivations for both the chain and the ladder are provided in Appendix~\ref{appe-JKT}. Remarkably, the spin-1/2 XY chain hosts a strictly flat zero-energy band. Furthermore, in the Kitaev spin ladder, the elementary spin excitations naturally fractionalize into itinerant Majorana fermions and static $Z_2$ gauge fluxes (visons), a phenomenon similar to the emergent fractionalization in the honeycomb lattice~\cite{kitaev_anyons_2006}. 

To investigate the charge dynamical properties of the in-gap states in the 1D and quasi-1D doped Kitaev-Hubbard models, we employ cluster perturbation theory (CPT). This numerical technique allows for the efficient calculation of zero-temperature single-particle spectral functions in strongly correlated electron systems~\cite{PhysRevB.66.075129,PhysRevB.85.144402,PhysRevB.94.125120,PhysRevLett.94.156404,PhysRevLett.107.010401,PhysRevLett.100.136402,PhysRevLett.107.010401,kohno_mott_2012,PhysRevLett.114.167201,science.abf5174,kmn8-y59j}. In this approach, the original infinite lattice is tiled with identical finite-size clusters, effectively forming a superlattice where each cluster serves as a unit cell. The model Hamiltonian is correspondingly decoupled into two parts: $H=H_c+H_T$. Here, $H_c$ encompasses the intra-cluster terms, including all local interactions and hoppings between intra-cluster sites, while $H_T$ contains the inter-cluster hopping terms. By treating $H_T$ perturbatively, the CPT single-particle Green's function is constructed via a Dyson-like equation:
\begin{equation}
G^{-1}_{\mathrm{cpt}}({\bm{Q}}, z)=G^{-1}_c(z)-V(\bm{Q}),
\end{equation}
where $\bm{Q}$ resides in the reduced Brillouin zone of the superlattice, $z$ is the complex frequency, and $G_c(z)$ represents the exact intra-cluster zero-temperature single-particle Green's function evaluated for $H_c$. In this paper, $G_c(z)$ is obtained utilizing the exact diagonalization (ED) method. $V(\bm{Q})$ is the Fourier transformed intercluster perturbation matrix, with matrix elements $V_{ab}(\bm{Q})=\sum_{\bm{R}}V_{ab}^{\bm{0} \bm{R}}e^{i\bm{Q}\cdot\bm{R}}$ ($V_{ab}^{\bm{0} \bm{R}}$ is the inter-cluster hopping amplitudes). Both $G_c$ and the inter-cluster hopping matrix $V(\bm{Q})$ are $2N_c \times 2N_c$ matrices, with $N_c$ being the number of sites per cluster. Because the cluster tiling explicitly breaks the translational symmetry of the original lattice, a periodization procedure is performed to recover the fully momentum-dependent Green's function:
\begin{equation}
G(\bm{k}, z)=\frac{1}{N_c} \sum_{a, b} \mathrm{e}^{-i\bm{k} \cdot\left(\bm{r}_a-\bm{r}_b\right)} G_{\mathrm{cpt}, a b}(\bm{Q}, z),
\label{CPTfinalG}
\end{equation}
where $\bm{k}$ is the wave vector defined within the first Brillouin zone (FBZ) of the original lattice and differs from $\bm{Q}$ by a reciprocal lattice vector $\bm{K}$ of the superlattice, i.e., $\bm{k} = \bm{Q} + \bm{K}$. $\bm{r}_a$ denotes the coordinate of lattice sites within the cluster. The retarded Green's function is obtained by performing an analytic continuation of the complex frequency: $G(\bm{k}, \omega) = G(\bm{k}, z \rightarrow \omega + i\eta)$, and we adopt a broadening factor of $\eta/t = 0.1$ unless otherwise specified. The spectral function is then extracted from the imaginary part of $G(\bm{k}, \omega)$: $A(\bm{k}, \omega)=-\frac{1}{\pi} \operatorname{Im} G(\bm{k}, \omega)$.

For the numerical computations in all three investigated lattice geometries, we employ $12$-site clusters. For the 1D XY chain, the specific cluster used is illustrated in Fig.~\ref{showclusters}(a). For the two-leg ladder Kitaev-Hubbard model, we utilize the cluster configuration depicted in Fig.~\ref{showclusters}(b). The many-body states with 12 electrons correspond to the half-filling case. And to investigate charge dynamics in doped system, our ED solver operates within a targeted Hilbert subspace of $11$ electrons, realizing a hole doping concentration of $\delta=1/12$. To prevent numerical singularities in our algorithm, specifically, to ensure that the application of a single-site creation or annihilation operator to the many-body ground state does not yield a null state, we introduce a slight deviation from the pure Kitaev point. Thus, our calculations are performed at $t^{\prime}=0.99$, which serves as an excellent asymptotic approximation to the exact Kitaev limit ($t^{\prime}=1$).
\begin{figure*}
\centering
\includegraphics[width=0.8\textwidth]{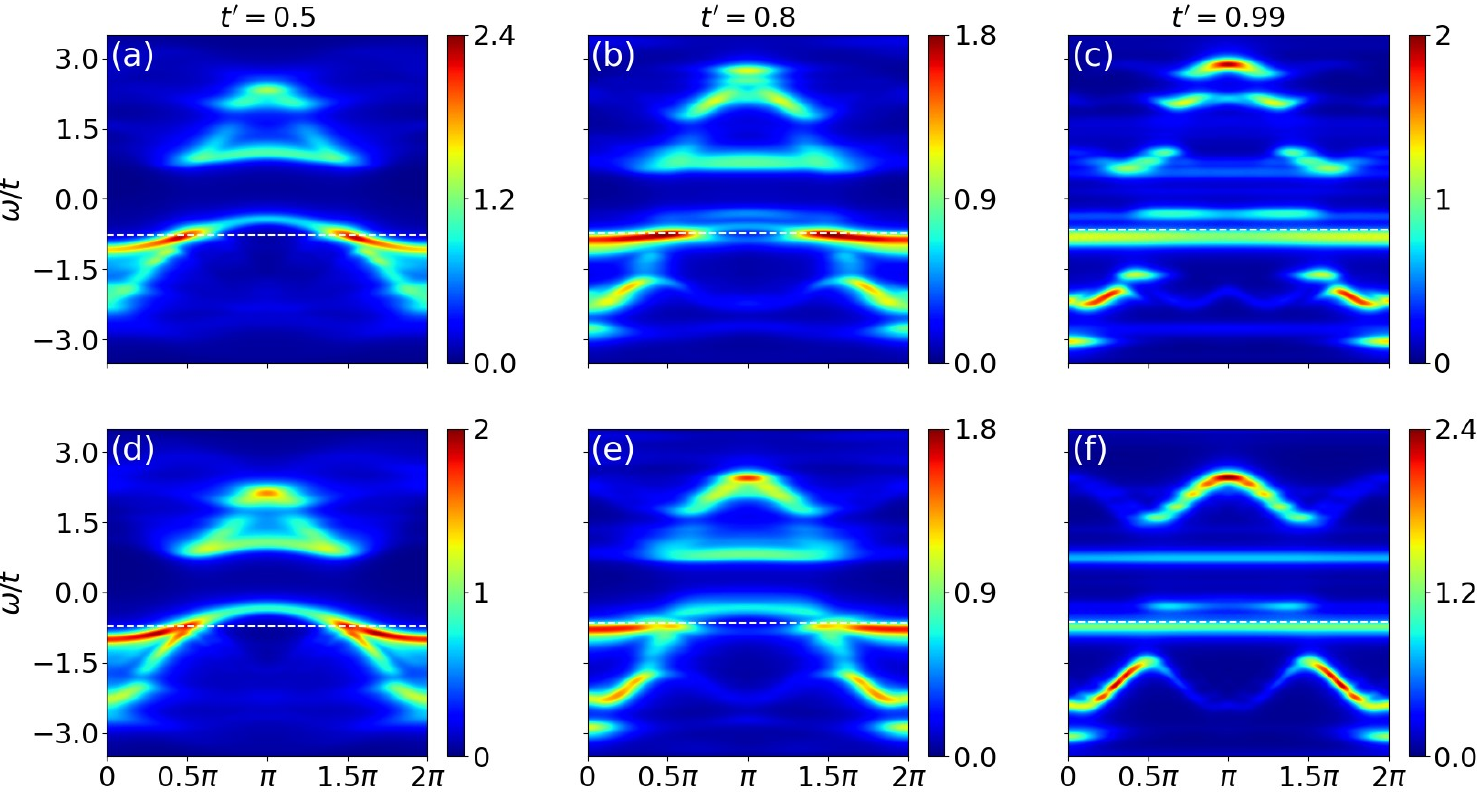}
\caption{Spectral functions of the Z chain ($U=3$). Panels (a)--(c) and (d)--(f) correspond to systems doped with a single spin-down and spin-up hole, respectively, for Kitaev-like hopping amplitudes $t^{\prime}=0.5, 0.8$, and $0.99$.}
\label{Zchain}
\end{figure*}

\section{Results}
\label{sec3}

\subsection{Z chain}
The single-particle spectrum of the pure Hubbard model ($t^{\prime}=0$) is extensively documented. Unlike conventional band insulators, where hole doping merely shifts the Fermi level within a rigid band structure, doping a Mott insulator dynamically generates new electronic states within the Mott gap, a phenomenon we detail via our CPT calculations in Appendix~\ref{appe-hubbard}. We now focus our studies to the 1D Z chain, which is characterized by additional $z$-type Kitaev-like hopping terms ($\sigma_{\alpha}=\sigma_z$). Because the hopping operator is diagonal in the spin basis, the spin orientation of the itinerant electrons is strictly preserved. Consequently, the total $S_z$ of the many-body state remains a conserved quantum number. Upon doping a single hole into the cluster, the many-body states corresponding to $S_z=1/2$ and $S_z=-1/2$ decouple into two disjoint Hilbert subspaces, which can be diagonalized independently.

Figure~\ref{Zchain} displays the systematic evolution of the spectral function for a single hole doped into the 12-site cluster as $t^{\prime}$ increases. The white dashed lines indicate the Fermi level, determined by integrating the spectral intensity. Figs~\ref{Zchain}(a)--(c) illustrate the spin-down hole doping sequence (6 spin-down electrons plus 5 spin-up electrons, $S_z=-1/2$). The higher-energy dispersive features of the spinon mode and the two holon modes---inherited from the pure Hubbard limit---gradually merge into a nearly dispersionless flat band. The most intriguing evolution occurs within the in-gap states: at $t^{\prime}=0.5$ [Fig.~\ref{Zchain}(a)], the in-gap band appears gapless around $k=\pi/2$, but as the Kitaev coupling strengthens to $t^{\prime}=0.8$ and $0.99$ [Figs.~\ref{Zchain}(b) and \ref{Zchain}(c)], a finite and distinct energy gap opens.

Figs.~\ref{Zchain}(d)--(f) present the corresponding spectra for a spin-up hole (5 spin-down electrons plus 6 spin-up electrons, $S_z=1/2$). Qualitatively, the spectral response to doping is largely symmetric with respect to the hole's spin orientation, featuring the formation of flat bands near the Fermi level and the emergence of gapped, relatively flat in-gap states at large $t^{\prime}$. The most notable spin-dependent asymmetry arises within the UHB at $t^{\prime}=0.99$: in the UHB, the spin-up hole [Fig.~\ref{Zchain}(f)] induces a well-defined flat band, whereas the spin-down hole [Fig.~\ref{Zchain}(c)] preserves a significant dispersive character. Nevertheless, the trajectory of the in-gap states is consistent across both spin sectors, transitioning from a gapless dispersion at weak $t^{\prime}$ to a strongly gapped, flattened band near the pure Kitaev limit.

\begin{figure}
\centering
\includegraphics[width=0.48\textwidth]{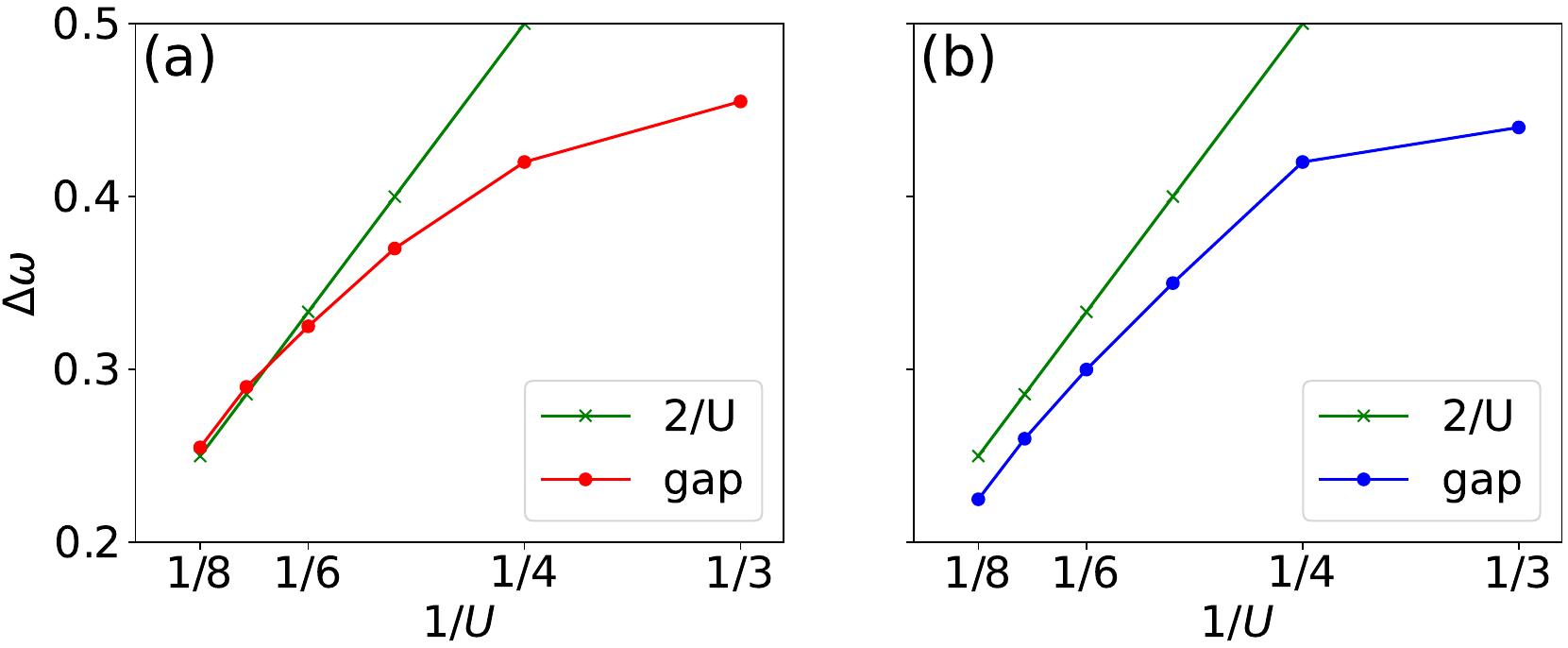}
\caption{Energy gap of the in-gap states for the single-hole-doped Z chain at $t^{\prime}=0.99$ for (a) a spin-down hole and (b) a spin-up hole, plotted as a function of the inverse interaction strength $1/U$. The green line indicates the spin gap of the Ising spin chain. The gap values are extracted from the frequency-resolved spectral intensity at momentum $k=\pi$.}
\label{Zgap}
\end{figure}
Formally, the relevance of the in-gap states to underlying spin excitations can be infered via the Lehmann representation of the single-particle spectral function, where the state created by injecting an electron ($c_{k \sigma}^{\dagger}| \mathrm{GS}\rangle$) maintains a non-zero overlap with the elementary magnon excitations of the half-filled background~\cite{PhysRevB.92.085129}. We find that a clear charge-spin correspondence  existed in the Kitaev limit ($t^{\prime}=1$). As discussed in Sec.~\ref{sec2}, the effective low-energy spin model in this limit maps onto the 1D antiferromagnetic Ising chain, where a localized single spin-flip excitation forms a strictly flat band with an energy penalty of $2t^{\prime 2}/U = 2/U$ [see Appendix~\ref{appe-XXZ}]. To quantitatively verify the correspondence found here, we extract the charge gap width $\Delta\omega$ by evaluating the energy difference between the spectral peaks corresponding to the top edge of the LHB and the in-gap state at a fixed momentum (e.g., $k=\pi$). Figs~\ref{Zgap}(a) and (b) illustrate the $U$-dependence of $\Delta\omega$ under spin-down and spin-up hole doping, respectively. As $U$ increases, the extracted charge gap $\Delta\omega$ scales inversely with $U$ and rapidly converges toward the theoretical spin-excitation gap of $2/U$ (indicated by the green lines). A subtle spin-dependent asymmetry is observed for large $U$: while the spin-up gap [Fig.~\ref{Zgap}(b)] precisely approaches the $2/U$ asymptotic line as expected, the spin-down gap [Fig.~\ref{Zgap}(a)] becomes slightly larger. The origin of this minor deviation can be understood as follows: in the Kitaev limit, the spin-down electrons lose their itinerancy completely, effectively acting as a static background potential $U$ that imposes an additional energy on the mobile spin-up electrons.

\subsection{XY chain}
\begin{figure}
\centering
\includegraphics[width=0.48\textwidth]{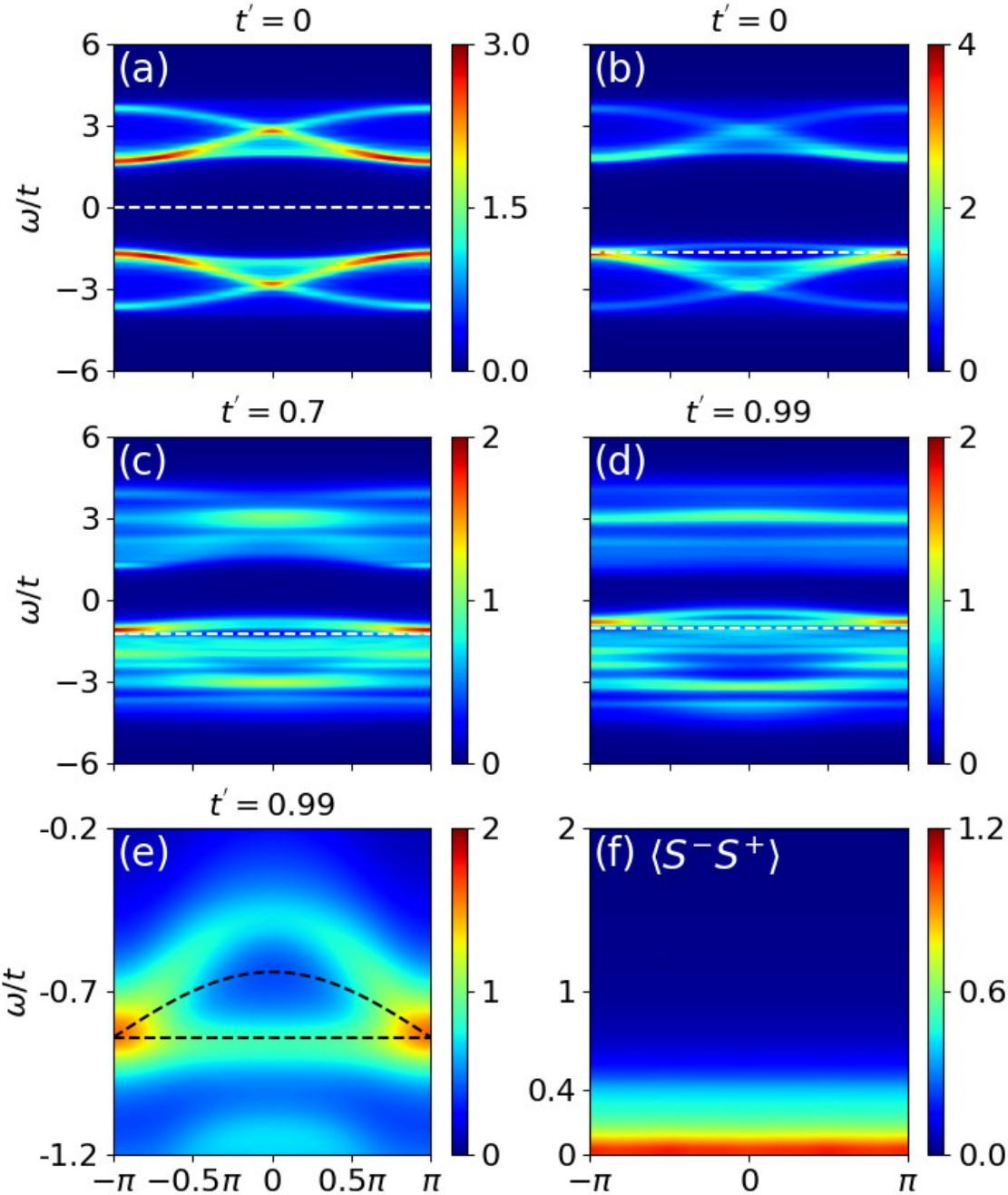}
\caption{Spectral functions of the XY chain ($U=5$). Panel (a) shows the half-filled system at $t^{\prime}=0$. Panels (b)--(d) display the single-hole-doped system for Kitaev-like hopping amplitudes $t^{\prime}=0.0$, $0.7$, and $0.99$, respectively. Panel (e) provides a magnified view of the energy window in (d) to clearly resolve the dispersion of the in-gap states. Panel (f) plots the transverse spin structure factor of the corresponding effective spin chain at $t^{\prime}=1$ ($J=0$, $K=2t^{\prime 2}/U=2/5$), calculated via ED on a 24-site cluster.}
\label{XYchain}
\end{figure}
The lattice geometry of the Kitaev-Hubbard model defined on the XY chain is illustrated in Fig.~\ref{showclusters}(a). The hopping interactions along the chain alternate between the two: one comprises conventional hopping combined with $x$-type Kitaev-like hopping (labeled $X$), while the other features conventional hopping combined with $y$-type Kitaev-like hopping (labeled $Y$). This intrinsic dimerization naturally halves the Brillouin zone and introduces a two-site unit cell with distinct two sublattices. Furthermore, the total $S_z$ is no longer a conserved quantum number due to $x$- and $y$-type hoppings. Our ED procedure is only restricted to the Hilbert subspace with conserved total particle number.

Figure~\ref{XYchain} details the evolution of the spectral function upon hole doping as a function of $t^{\prime}$ at a fixed $U=5$. We first note that for the half-filled case at $t^{\prime}=0$ [Fig.~\ref{XYchain}(a)], the artificial introduction of the two-sublattice unit cell yields a standard band-folding phenomenon compared to the pure Hubbard chain shown in Fig.~\ref{pureHubbardchain}(a). Upon doping a single hole, the emergent in-gap states naturally extend across the entire reduced Brillouin zone, as depicted in Fig.~\ref{XYchain}(b). As $t^{\prime}$ is progressively tuned from 0 to 0.99 [Figs.~\ref{XYchain}(b)--(d)], the well-defined dispersive modes within both the UHB and LHB gradually lose their momentum-dependent characteristics, smearing into broad spectral continua. Most crucially, we focus on the trajectory of the in-gap states residing immediately above the Fermi level. At $t^{\prime}=0.0$ [Fig.~\ref{XYchain}(b)], these states simply reflect the folded dispersion of the pure Hubbard chain. At the intermediate coupling $t^{\prime}=0.7$ [Fig.~\ref{XYchain}(c)], the single in-gap mode exhibits precursors of splitting into two distinct branches, though they are not yet fully resolved. Approaching the Kitaev limit at $t^{\prime}=0.99$ [Fig.~\ref{XYchain}(d)], this splitting becomes definitively clear: the in-gap states separate into a dispersive upper branch and an approximately flat lower branch that is pinned closely to the Fermi level. 

This remarkable structure can be understood through its low-energy effective spin model. At the pure Kitaev limit ($t^{\prime}=1$), the strongly coupled XY chain maps onto the 1D $S=1/2$ XY Kitaev spin chain. As detailed in Appendix~\ref{appe-JKT}, this spin model can be analytically diagonalized via the Jordan-Wigner transformation, yielding a fermionic spectrum composed of a strictly zero-energy flat band and a dispersive band at $K_x=K_y$. Remarkably, the analytical non-interacting spectrum matches the kinetic dispersion of the in-gap states qualitatively. This correspondence is highlighted in Fig.~\ref{XYchain}(e), a magnified low-energy window of Panel (d), where the black dashed lines trace the analytical dispersion given by Eq.~(\ref{XYdispersion}) utilizing the effective coupling $K=2t^{\prime 2}/U=2/5$ ($S_i^{\alpha}S_{i+1}^{\alpha}={\sigma}_i^{\alpha}{\sigma}_{i+1}^{\alpha}/4$, $K_x=K_y=0.1$). Quantitatively, the energy elevation of the numerically observed in-gap states relative to the analytical black dashed lines is a natural consequence of finite-$U$ effects. To further solidify this connection, Fig.~\ref{XYchain}(f) displays the transverse dynamic spin structure factor of the effective spin chain, computed via ED on a 24-site ring with periodic boundary condition. The presented data specifically captures intra-sublattice spin correlations ($S_i^-S_j^+$ where $i$ and $j$ belong to the same sublattice). Due to the non-zero spin correlations only within NN bonds in Kitaev physics~\cite{PhysRevLett.112.207203}, these excitations are completely momentum-independent, manifesting as a broad, gapless continuum. We note a slight energy offset in Fig.~\ref{XYchain}(d), where the Fermi level sits marginally below the nominally "zero-energy" flat branch of the in-gap states; This minor deviation originates from the numerical calibration of the Fermi level: because the spectral intensity is integrated over a finite frequency window rather than extending strictly to $\omega \in (-\infty,\infty)$, a truncation error is introduced. The striking correspondence between the in-gap states and the fermionic quasiparticles of spin chain provides evidence that the kinetic dispersion of the electronic states is driven by emergent quasiparticles. The broad continuum observed in the spin excitations reflects the fractionalization of local $S=1$ spin flips into itinerant Majorana fermions. Because both the charge and spin excitations are ultimately constructed from these same fractionalized elementary quasiparticles, their dynamical signatures are profoundly intertwined.

\subsection{The two-leg ladder}
\begin{figure}
\centering
\includegraphics[width=0.48\textwidth]{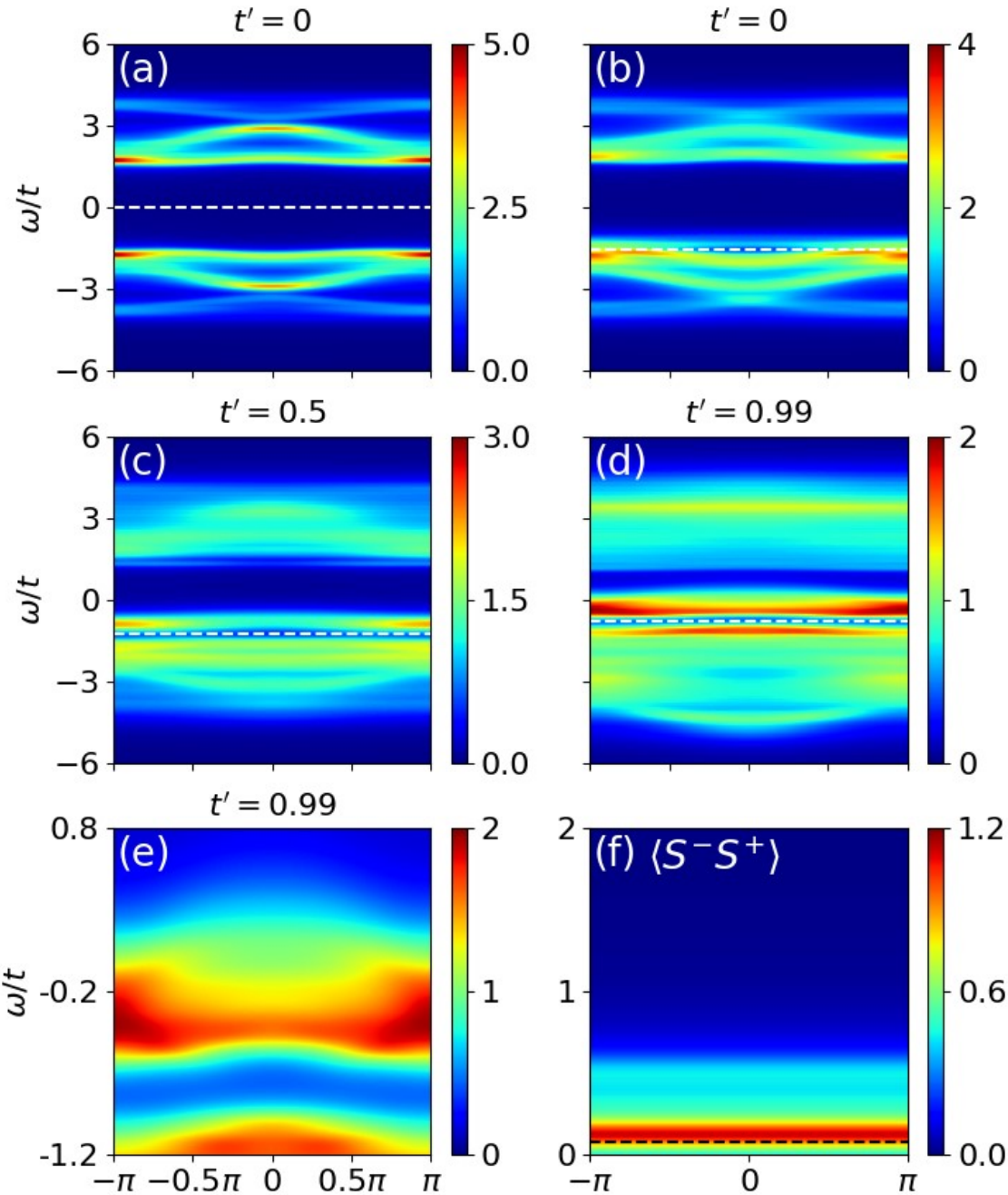}
\caption{Spectral functions of the two-leg ladder ($U=5$). Panel (a) shows the half-filled system at $t^{\prime}=0$. Panels (b)--(d) display the spectra for the single-hole-doped system with $t^{\prime}=0.0$, $0.5$, and $0.99$, respectively. Panel (e) provides a magnified view of the energy window in (d) to resolve the continuum nature of the in-gap states. Panel (f) plots the transverse spin structure factor of the corresponding effective spin ladder at $t^{\prime}=1$ ($J=0$, $K=2t^{\prime 2}/U=2/5$), calculated via ED on a $2 \times12$ cluster.}
\label{XYZladder}
\end{figure}
The specific two-leg ladder cluster employed in our CPT calculations is illustrated in Fig.~\ref{showclusters}(b). By imposing periodic boundary condition along the rung direction, this double-chain geometry becomes topologically equivalent to the square-lattice ladder depicted in Fig.~\ref{showclusters}(c). Notably, the real-space periodicity along the rung direction is exactly twice the NN bond length, yielding a single unit cell along the rung, thereby locking the conserved rung direction momentum at $k_{2} = 0$ and effectively rendering the system a quasi-1D model.

Figure~\ref{XYZladder}(a) establishes the baseline spectral function for the half-filled ladder at $U=5$ in the absence of Kitaev-like hopping ($t^{\prime}=0$). Upon doping the cluster with a single hole, emergent in-gap states immediately appear above the LHB, as shown in Fig.~\ref{XYZladder}(b). The subsequent sequence in Figs.~\ref{XYZladder}(b)--(d) denotes the spectral evolution with the Kitaev-like hopping $t^{\prime}$ from $0.0$, to $0.5$, and $0.99$, respectively. With increasing $t^{\prime}$, the in-gap states differs significantly from the 1D scenario. At intermediate coupling ($t^{\prime}=0.5$) [Fig.~\ref{XYZladder}(c)], a distinct energy gap fully separates the in-gap states from the bulk LHB continuum. This isolation persists up to the near-Kitaev limit ($t^{\prime}=0.99$), where the spectral weight of the in-gap states itself becomes highly diffuse, forming a broad continuum. This continuous character is unequivocally resolved in the magnified low-energy window presented in Fig.~\ref{XYZladder}(e).

In the pure Kitaev limit ($t^{\prime}=1.0$), the low-energy effective Hamiltonian maps onto the Kitaev spin ladder. As detailed in Appendix~\ref{appe-JKT}, this ladder model, much like its 2D honeycomb counterpart, is exactly solvable by mapping to itinerant Majorana fermions coupled to static $Z_2$ gauge fields. A fundamental consequence of this mapping is the decomposition of the Hilbert space into distinct gauge-flux sectors, supporting localized, gapped vison excitations above the ground state~\cite{PhysRevLett.98.247201,PhysRevLett.112.207203,PhysRevB.108.045151}. Figure~\ref{XYZladder}(f) displays the transverse dynamic spin structure factor of this effective spin ladder under periodic boundary condition, aligning perfectly with prior calculations~\cite{PhysRevB.99.224418}. The spin excitations robustly manifest as a gapped continuum---a hallmark signature of fractionalized quasiparticles. The black dashed line explicitly demarcates the minimum energy required to excite a vison pair by analytical calculations [see Appendix~\ref{appe-JKT}]. Returning to our CPT results, the observation that the in-gap states form a continuous spectrum rather than a set of coherent dispersive bands is a striking and profound phenomenon. We argue that these in-gap states directly inherit the fractionalized nature of the underlying Kitaev spin liquid~\cite{PhysRevB.99.195112,PhysRevB.99.224418}; the charge excitations dynamically fractionalize, inheriting the nature of Majorana fermions and visons. And we conclude that the persistent energy gap separating the in-gap states from the Fermi level in Fig.~\ref{XYZladder}(d) is a direct spectroscopic manifestation of the $Z_2$ vison gap inherent to the ladder lattice.

\section{conclusion}
\label{sec4}
In summary, we have investigated the fundamental interplay between charge dynamics and spin excitations in doped Mott insulators. And we established a definitive correspondence between doping-induced in-gap states and the underlying spin physics. Through selectively tuning spin anisotropies via Kitaev-like hopping of the Kitaev-Hubbard model in the strong-coupling limit, our calculations demonstrate how the kinetic dispersion of charge states maps onto exotic spin dynamics across distinct lattice geometries. Specifically, in the Z chain, we observed that the in-gap states flatten and open a characteristic charge gap scaling as $2t^{\prime 2}/U$, identically tracking the spin-gap behavior of the Heisenberg-to-Ising crossover. In the XY chain, the charge dispersion replicates the free-fermion spectrum of the corresponding spin chain. Strikingly, extending this to the two-leg ladder, the in-gap states dissolve into a broad spectral continuum featuring a distinct energy gap, providing a direct charge signature of gapped $Z_2$ visons inherent to the Kitaev spin liquid. The physical picture emerging from these models is that the in-gap charge states and the $S=1$ spin excitations are not merely phenomenologically correlated; rather, they are fundamentally composed of the same emergent quasiparticles. Upon doping, the induced charge excitations dynamically fractionalize, coupling deeply with the background itinerant Majorana fermions and static gauge fluxes (visons). It is this shared quasiparticle composition that determines their profoundly intertwined dynamics, leading the in-gap states to spectrally mirror the underlying spin behavior. Our findings demonstrate that in-gap states can serve as a sensitive spectroscopic probe for spin excitations, offering a new pathway to probe the fractionalized continuum in spin liquids. Thus, exotic spin dynamics may be accessible through standard charge spectroscopies like ARPES and STM.


\begin{acknowledgments}
We gratefully acknowledge discussions with Wei Wang, Zhao-Long Gu and Li-Wei He. This work was supported by National Key Projects for Research and Development of China (No. 2021YFA1400400 and No. 2024YFA1408104), National Natural Science Foundation of China (No. 12374137, No. 12434005, and No. 12550405).
\end{acknowledgments}

\appendix

\section{Dynamic spin structure factor of 1D XXZ spin chain}
\label{appe-XXZ}
\begin{figure}
\centering
\includegraphics[width=0.48\textwidth]{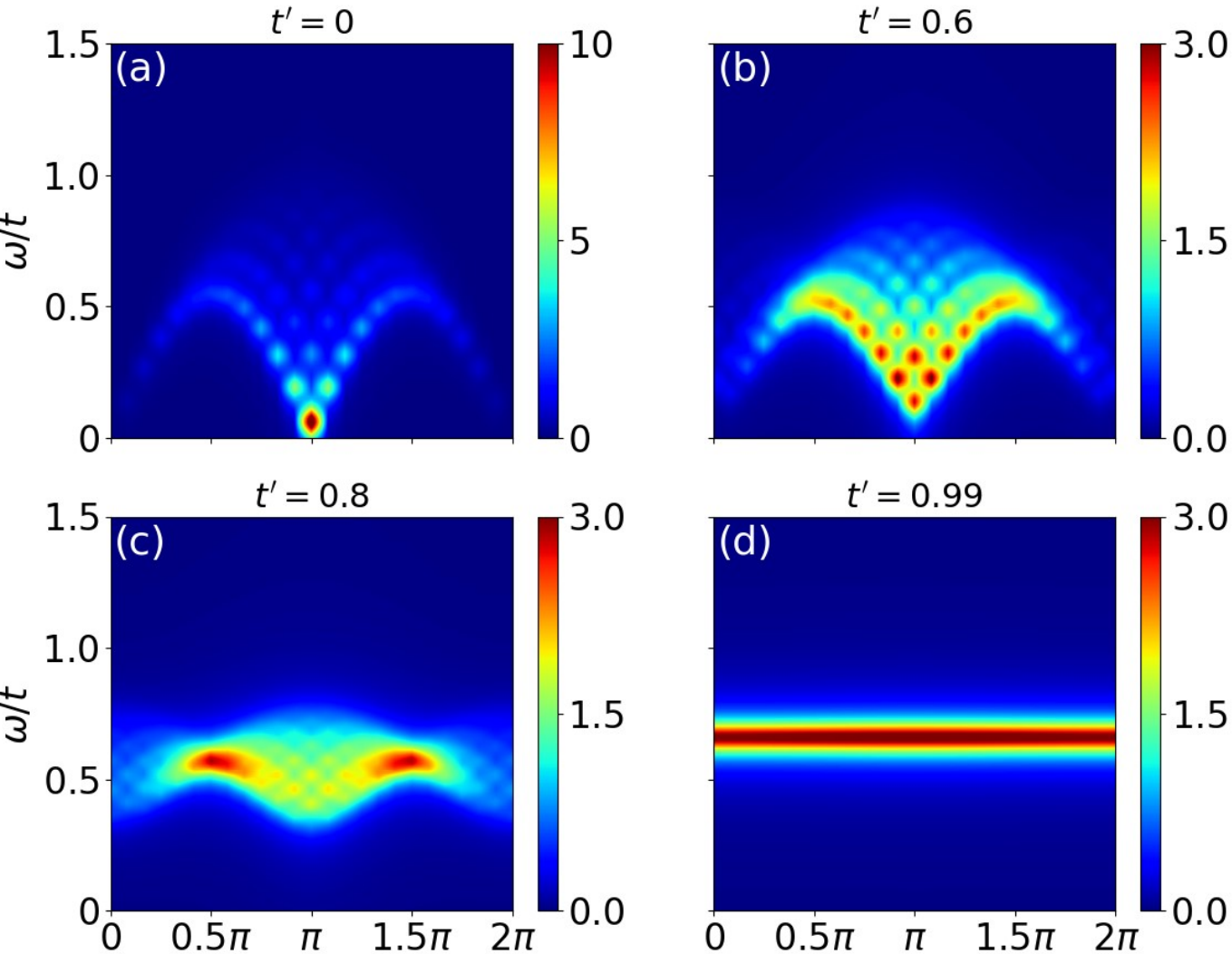}
\caption{Dynamic transverse spin structure factor of the 1D XXZ spin chain, obtained via ED on a 24-site chain with a broadening factor $\eta=0.05$. The effective exchange parameters are $J_x=J_y=(1-t^{\prime 2})/U$ and $J_z=(1+t^{\prime 2})/U$, with $U=3$. (a) $t^{\prime}=0$, recovering the isotropic 1D antiferromagnetic Heisenberg model. (b) $t^{\prime}=0.6$. (c) $t^{\prime}=0.8$. (d) $t^{\prime}=0.99$, which closely approximates the 1D antiferromagnetic Ising model limit.}
\label{hkspin}
\end{figure}
To better elucidate the correspondence between the dispersion of the in-gap states and the spin excitations in the case of 1D Kitaev-Hubbard Z chain, we analyze its low-energy effective spin model in the strong-coupling limit ($U \gg t, t^{\prime}$), which maps onto the well-known 1D XXZ spin model. Driven by the methodological advancements from the Bethe ansatz to quantum affine group techniques, the exact analytical solutions of the 1D XXZ spin chain have been systematically expanded from ground-state properties to dynamic spin structure factors~\cite{PhysRev.112.309,PhysRev.150.321,PhysRev.150.327,PhysRev.151.258,Jimbo1994AlgebraicAO,Michio_Jimbo_1996,PhysRevB.57.11429,PhysRevLett.96.257202,KITANINE2000554,PhysRevB.54.R12669}. Here, we employ ED to compute the transverse dynamic spin structure factor of a 24-site chain with periodic boundary condition. Consequently, the momentum resolution is restricted to 24 discrete points, inevitably introducing finite-size effects into the spectral features. The resulting spectra are presented in Fig.~\ref{hkspin}, labeled by the Kitaev-like hopping amplitude $t^{\prime}$ of the original electronic model, with effective spin couplings determined by Eq.~(\ref{effhamiltonian}) at $U=3$. As depicted in Fig.~\ref{hkspin}(a) for the isotropic limit ($t^{\prime}=0$), the system reduces to the 1D antiferromagnetic Heisenberg model in which the spin excitation is gapless in the thermodynamic limit~\cite{PhysRev.128.2131}. In this case,  the $S=1$ magnetic excitation consists of a pair of fractionalized $S=1/2$ spinons, and the lower boundary of the observed continuum delineates the single-spinon dispersion. Figures~\ref{hkspin}(b) and \ref{hkspin}(c) display the spectra at $t^{\prime}=0.6$ and $0.7$, respectively. In accordance with previous theoretical findings that any infinitesimal Ising-like anisotropy generates a gap~\cite{PhysRevB.57.11429}, a pronounced spin excitation gap is clearly resolved in our numerical results. When approaching the pure Kitaev limit ($t^{\prime}=0.99$) [Fig.~\ref{hkspin}(d)], the Heisenberg exchange is severely suppressed, and the system is well-approximated by the 1D antiferromagnetic Ising chain. Here, the $S=1$ magnetic excitation manifests as a nearly dispersionless flat band centered at an energy cost of roughly $2t^{\prime 2}/U \approx 2/3$. 

\section{Analytical solution of $S=1/2$ Kitaev spin chain and spin ladder by Jordan-Wigner transformation}
\label{appe-JKT}
\begin{figure}
\centering
\includegraphics[width=0.48\textwidth]{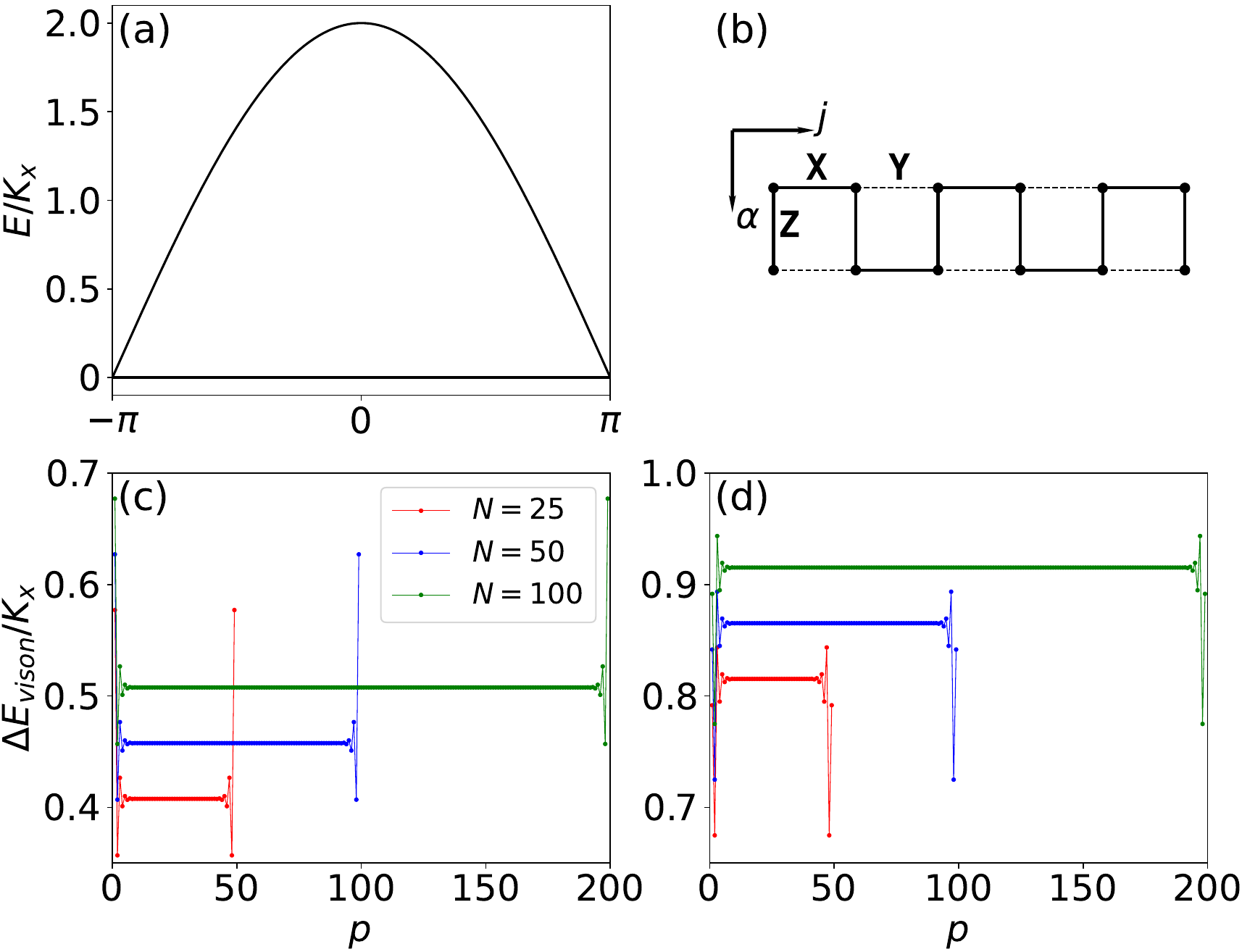}
\caption{(a) The fermionic dispersion of the 1D $S=1/2$ Kitaev spin chain ($K_x=K_y=1$). (b) Schematic representation of the two-leg Kitaev spin ladder. The solid line defines the 1D ordering of lattice sites adopted for the Jordan-Wigner transformation. (c)  Energy cost of a single $Z_2$ gauge flux (vison) as a function of spatial position under open boundary condition ($K_x=K_y=K_z=1$). Data for system sizes $N=50$ and $100$ are vertically offset by $0.05$ and $0.10$, respectively, for clarity. In the bulk, the single-vison energy is independent of its position. (d) Energy of a vison pair as a function of spatial separation under periodic boundary condition, with one vison spatially pinned at the boundary plaquette.}
\label{visongap}
\end{figure}

The 1D XY chain and the two-leg ladder geometries studied in the main text are obtained from the honeycomb lattice. Single and double chains extracted from the $S=1/2$ Kitaev model therefore are the low-energy effective spin models of these quasi-1D Kitaev-Hubbard model in the strong-coupling limit, which can be analytically diagonalized via the Jordan-Wigner transformation, mapping them exactly to free-fermion systems~\cite{PhysRevLett.98.087204}. Let us first consider the 1D XY chain. Assuming a chain comprising $N$ unit cells (equivalent to $2N$ sites as there are two sublattices), the spin Hamiltonian can be expressed as a sum over one sublattice:
\begin{equation}
H=\sum_{j=2 m-1} \left( K_x \sigma_j^x \sigma_{j+1}^x+K_y \sigma_{j+1}^y \sigma_{j+2}^y \right),
\end{equation}
where $m \in [1, N]$ and is an integer. By the Jordan-Wigner transformation:
\begin{equation}
\begin{aligned}
\sigma_j^x & =\left(a_j^{\dagger}+a_j\right) e^{i \pi \sum_{i<j} a_i^{\dagger} a_i}, \\
\sigma_j^y & =-i\left(a_j^{\dagger}-a_j\right) e^{i \pi \sum_{i<j} a_i^{\dagger} a_i}, \\
\sigma_j^z & =2 a_j^{\dagger} a_j-1, 
\end{aligned}
\end{equation}  
and a sublattice-dependent Majorana representation of the spinless fermion $a$:
\begin{equation}
\begin{array}{ll}
c_j=i\left(a_j^{\dagger}-a_j\right), \quad d_j=a_j^{\dagger}+a_j, & j=2 m-1, \\
c_j=a_j^{\dagger}+a_j, \quad d_j=i\left(a_j^{\dagger}-a_j\right), & j=2 m, 
\end{array}
\label{majoranarep}
\end{equation} 
where $c_j$ and $d_j$ are Majorana operators. In this representation, the Hamiltonian simplifies to:
\begin{equation}
H=-i \sum_{j=2 m-1}\left(K_x c_j c_{j+1}-K_y c_{j+1} c_{j+2}\right).
\end{equation}    
Remarkably, the Hamiltonian solely involves the itinerant $c$ Majorana operators, completely decoupling from the static $d$ operators. Since each operator $i d_j d_{j+1}$ possesses eigenvalues $\pm 1$, this decoupling gives rise to a massive $2^N$-fold macroscopic degeneracy in the energy spectrum. By recombining the static $d$ Majoranas into complex Dirac fermions, $d^f_m = (d_{2m-1} + i d_{2m})/2$, one obtains a strictly zero-energy flat band across the entire Brillouin zone. To determine the dispersive energy bands governed by the $c$ operators, we impose periodic boundary condition and perform a Fourier transform (take the length of the translational vector as unit $1$) to momentum space. This casts the Hamiltonian into a bilinear matrix form:
\begin{equation}
H=-i \sum_k C^{\dagger}\left(\begin{array}{cc}
0 & \alpha \\
-\alpha^* & 0
\end{array}\right) C,
\end{equation}    
where $\alpha=\left(K_x e^{-i k/2}+K_y e^{i k/2}\right)$, and the spinor is defined as
\begin{equation}
C=\left(\begin{array}{l}
c_{-k, A} \\
c_{-k, B}
\end{array}\right).
\end{equation}
Diagonalization of this $2 \times 2$ matrix yields the eigenvalues:
\begin{equation}
E(k)= \pm \sqrt{|\alpha(k)|^2}= \pm \sqrt{K_x^2+K_y^2+2 K_x K_y \cos (k)}.
\label{XYdispersion}
\end{equation}
This spectrum inherently contains a particle-hole redundancy, $E(-k) = -E(k)$. The physical fermionic quasiparticle dispersion is strictly given by the positive energy branches. For $K_x = K_y = 1$, the resulting eigenvalue spectrum, comprising both the dispersive $c$-band and the flat $d$-band, is plotted in Fig.~\ref{visongap}(a). Notably, this analytical non-interacting fermionic quasiparticle dispersion perfectly matches the kinetic profile of the hole-doping-induced in-gap states previously calculated for the interacting Kitaev-Hubbard electronic XY chain.

Next, we turn to the two-leg spin ladder, depicted in Fig.~\ref{visongap}(b) with $N=3$ unit cells. We denote the chain index by $\alpha = 1, 2$ and label the sites within each chain sequentially by index $j \in [1, 2N]$. The ladder Hamiltonian is naturally decomposed into intra-chain ($H_1, H_2$) and inter-chain ($H_I$) interactions: 
\begin{equation}
\begin{aligned}
&H=H_1+H_2+H_I,\\
&\begin{aligned}
& H_1=\sum_{j=2 m-1} K_x \sigma_{j, 1}^x \sigma_{j+1,1}^x+K_y \sigma_{j+1,1}^y \sigma_{j+2,1}^y, \\
& H_2=\sum_{j=2 m-1} K_x \sigma_{j+1,2}^x \sigma_{j+2,2}^x+K_y \sigma_{j, 2}^y \sigma_{j+1,2}^y, \\
& H_I=\sum_{j=2 m-1} K_z \sigma_{j, 1}^z \sigma_{j, 2}^z+K_z \sigma_{j+1,1}^z \sigma_{j+1,2}^z.
\end{aligned}
\end{aligned}
\end{equation}
To implement the Jordan-Wigner transformation on this quasi-1D geometry, we must define a strict 1D ordering of the lattice sites, delineated by the solid path in Fig.~\ref{visongap}(b). Following this sequence from left to right, the $4N$ spins are transformed into spinless fermions. Then, by adopting a sublattice-dependent Majorana basis analogous to the 1D chain, the Majorana Hamiltonian for the ladder lattice reads:
\begin{equation}
\begin{aligned}
H= -i \sum_{j=2 m-1} &\left[K_x c_{j, 1} c_{j+1,1}-K_y c_{j+1,1} c_{j+2,1}\right. \\
&+K_x c_{j+1,2} c_{j+2,2}-K_y c_{j, 2} c_{j+1,2} \\
&+K_z D_{j} c_{j, 1} c_{j, 2}+K_z D_{j+1} c_{j+1,1} c_{j+1,2}].
\end{aligned}
\label{laddermajorana}
\end{equation} 
Here, the rung operators $D_{j}=-i d_{j, 1} d_{j, 2}$ explicitly commute with the Hamiltonian ($[H, D_j] = 0$) and possess eigenvalues $D_j = \pm 1$. Consequently, akin to the 2D honeycomb Kitaev model, $D_{j}$ acts as a static $Z_2$ gauge field, reducing the problem to free itinerant Majorana fermions coupled to a static gauge background. According to Lieb's theorem~\cite{PhysRevLett.73.2158}, the ground state of a bipartite ladder at half-filling is stabilized by a $\pi$-flux configuration in every square plaquette~\cite{PhysRevB.86.205412,PhysRevB.96.205109}. In the parameterization of Eq.~(\ref{laddermajorana}), the gauge fields on the $xy$-type bonds are inherently fixed. To realize the $\pi$-flux ground state, we simply fix all $z$-type bond gauge fields uniformly to $D_{j}=1$. In this $D_j=1$ gauge, translational symmetry along the chain is restored and physical fermionic quasiparticle dispersion in momentum space can also be obtained. However, beyond the itinerant fermions, localized $Z_2$ gauge flux excitations (visons) exist in the ladder. The energy cost of a single vison is defined as the energy difference between a specific single-vortex gauge configuration and the uniform $\pi$-flux ground state. This excitation gap is calculated by writing the Majorana Hamiltonian for a given gauge configuration as $H(A)= -i \sum_{j, k} A_{j k} c_j c_k$, where $A$ is a real antisymmetric matrix dictated by the spatial distribution of the gauge fields. By applying an orthogonal transformation $Q \in O(4N)$ to map to a diagonal Majorana basis $\left(b_1, \ldots, b_{4 N}\right) = \left(c_1, \ldots, c_{4 N}\right) Q$, the matrix $A$ is block-diagonalized with skew-symmetric blocks containing the eigenvalues $\pm i\varepsilon_m$:
\begin{equation}
A=Q\left(\begin{array}{ccccc}
0 & \varepsilon_1 & & & \\
-\varepsilon_1 & 0 & & & \\
& & \ddots & & \\
& & & 0 & \varepsilon_{2N} \\
& & & -\varepsilon_{2N} & 0
\end{array}\right) Q^{\mathrm{T}}.
\end{equation}
Recombining the $b$ operators into complex Dirac fermions $f_m = \frac{1}{2} (b_{2m-1} + i b_{2m})$, the Hamiltonian is fully diagonalized: $H= -4 \sum_{m=1}^{2N} \varepsilon_m (f_m^{\dagger} f_m- 1/2)$. The ground state energy for a given gauge sector is therefore $E_{GS}=-2 \sum_{m=1}^{2N} |\varepsilon_m| = - \operatorname{Tr}|i A|$. 

For $K_x=K_y=K_z=1$, the spatially resolved energy cost of a single vison for open finite ladders ($N=25, 50, 100$) is plotted in Fig.~\ref{visongap}(c). Excluding finite-size boundary effects, the bulk vison gap is robustly independent of spatial location, saturating to a finite value slightly above 0.4. Furthermore, under periodic boundary condition, topological constraints dictate that vison excitations must be created in pairs. Fig.~\ref{visongap}(d) illustrates the excitation energy of such a vison pair as a function of their spatial separation, where one vison is pinned at the boundary plaquette while the other is moved across the lattice. Crucially, as the distance between the two visons increases, the total energy converges to a flat plateau that is twice the single-vison energy in Fig.~\ref{visongap}(c). This numerical correspondence highlights that visons behave as independent quasiparticles. The noticeable deviation from this plateau only at small separations unambiguously confirms that the mutual interaction between visons is short-ranged.

\section{Spectral function of 1D Hubbard model}
\label{appe-hubbard}
\begin{figure}
\centering
\includegraphics[width=0.48\textwidth]{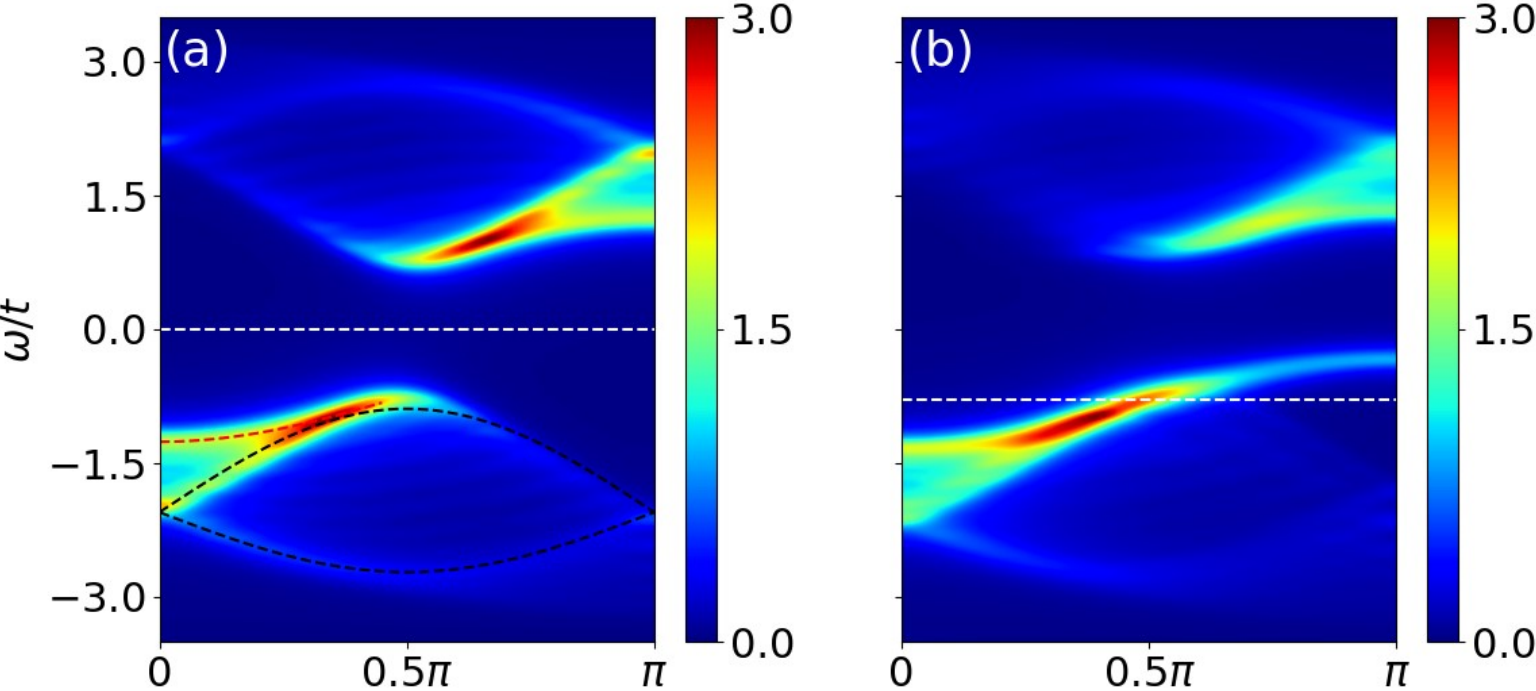}
\caption{Spectral functions of the 1D Hubbard model ($t^{\prime}=0$, $U=3$) for (a) the half-filled system and (b) the system doped with a single hole. The spectrum in (b) is independent of the spin orientation of the doped hole.}
\label{pureHubbardchain}
\end{figure}
We calculated the zero-temperature single-particle spectral function of the pure 1D Hubbard model (without $t^{\prime}$) via CPT, depicted in Fig.~\ref{pureHubbardchain}(a) for half filling, which is in excellent agreement with previous studies~\cite{Kohno_2018,PhysRevLett.92.256401,he_effect_2022}. The spectrum is dominated by well-defined UHB and LHB, separated by a robust Mott gap, with both bands exhibiting particle-hole symmetry. Notably, the 1D Hubbard model is a paradigmatic integrable system with an exact analytical solution via the Bethe ansatz~\cite{PhysRevLett.20.1445}. Based on the rigorous physical picture provided by this method, the broad continuum features and the highly structured internal branches within the LHB can be understood through fractionalized elementary excitations, namely spinons, holons, antiholons, and string solutions~\cite{kohno_spectral_2010,takahashi1972one}. Specifically, three primary excitation modes within the LHB can be clearly identified, each named primarily for the specific fractionalized quasiparticle that governs its momentum dispersion. The prominent dispersive branch indicated by the red dashed line, extending over the momentum interval $k \in [0, \pi/2]$, is recognized as the spinon mode. Situated at slightly lower energies is a mode that merges with the spinon branch near $k = \pi/2$. This feature originates from a composite excitation involving a holon, a spinon, and an antiholon. Crucially, in this composite state, the momenta of the spinon and antiholon are kinematically pinned at fixed values. Consequently, the momentum dispersion of this branch is entirely dictated by the holon kinematics, and thus it is conventionally referred to as the holon mode. Finally, the low-energy branch with relatively weak spectral weight is similarly driven by holon dynamics and is recognized in the literature as the shadow band~\cite{PhysRevLett.77.1390}. The two holon-governed modes are indicated by black dashed lines in Fig.~\ref{pureHubbardchain}(a). Figure~\ref{pureHubbardchain}(b) illustrates the spectral function upon the introduction of a single doped hole. While the overarching Hubbard band struture remains largely unchanged compared to the half-filled case, the fundamental signature of doping a Mott insulator manifests via a stark spectral weight transfer. Most notably, a distinct new dispersive branch emerges and resides within the Mott gap. These newly formed states, isolated from the Hubbard band, correspond to the so-called in-gap states.


\bibliography{main}

\end{document}